\documentclass[12pt]{article}

\newcommand{\be}{\begin{equation}}
\newcommand{\ee}{\end{equation}}
\newcommand{\bi}[1]{\vspace{-3mm} \bibitem{#1}}

\begin{document}

\begin{center}
Chaos Vol.15. No.2. (2005) 023102
\end{center}

\begin{center}
{\large \bf Fractional Fokker-Planck Equation for Fractal Media}
\vskip 5 mm

{\large \bf Vasily E. Tarasov } \\

\vskip 3mm
{\it Skobeltsyn Institute of Nuclear Physics, \\
Moscow State University, Moscow 119992, Russia}

E-mail: tarasov@theory.sinp.msu.ru
\end{center}
\vskip 11 mm

\begin{abstract}
We consider the fractional generalizations of 
equation that defines the medium mass.
We prove that the fractional integrals can be used to 
describe the media with noninteger mass dimensions.
Using fractional integrals, we derive the fractional generalization 
of the Chapman-Kolmogorov equation (Smolukhovski equation).  
In this paper fractional Fokker-Planck equation for fractal media
is derived from the fractional Chapman-Kolmogorov equation. 
Using the Fourier transform, we get the
Fokker-Planck-Zaslavsky equations that have fractional coordinate  derivatives.
The Fokker-Planck equation for the fractal media is an equation with fractional
derivatives in the dual space. 
\end{abstract}

\vskip 7mm


{\bf 
The application of fractals in physics is far ranging, 
from the dynamics of fluids in porous media 
to resistivity networks in electronics. 
The cornerstone of fractals is the meaning of dimension, 
specifically the fractal dimension. 
Fractal dimension can be best calculated by box counting 
methods which means drawing a box of size R 
and counting the mass inside. 
Fractal models of media are enjoying considerable
popularity. This is due in part to the relatively 
small number of parameters that can define a 
fractal medium of great complexity and rich structure.
Derivatives and integrals of fractional order have found many
applications in recent studies of scaling phenomena.
It is interesting to use fractional integration to consider the 
fractional generalizations of the equations that describe fractal media.
We use the fractional integrals to describe fractal media 
with noninteger mass dimensions.
The fractional integrals can be used not only to calculate 
the mass dimensions of fractal media. 
Fractional integration can be used
to describe the dynamical processes in the fractal media.
Using fractional integrals, we derive the fractional generalization 
of the Chapman-Kolmogorov equation (Smolukhovski equation).  
The fractional Fokker-Planck equation for fractal media is derived from
the fractional Chapman-Kolmogorov equation. 
Using the Fourier transform, we get the Fokker-Planck-Zaslavsky equations 
that have coordinate fractional derivatives.

}

\section{Introduction}

The application of fractals in physics \cite{It,Zas2,Zas3,Zas4}
is far ranging, from the dynamics of fluids in porous media 
to resistivity networks in electronics. 
Derivatives and integrals of fractional order have found many
applications in recent studies of scaling phenomena 
\cite{Zas2,Zas3,1,2,3,4,Zas}.

The natural questions arise: What could be the physical
meaning of the fractional integration?
This physical meaning can be following:
the fractional integration can be considered as an
integration in some fractional space.
In order to use this interpretation we must define
a fractional space.
The first interpretation of the fractional space
is connected with fractional dimension space.
The fractional dimension interpretation follows from the
formulas for dimensional regularizations.
If we use the well-known formulas for dimensional 
regularizations \cite{Col}, then we get
that the fractional integration can be considered as an 
integration in the fractional dimension space \cite{chaos}
up to the numerical factor
$\Gamma(\alpha/2) /[ 2 \pi^{\alpha/2} \Gamma(\alpha)]$.
This interpretation was suggested in Ref. \cite{chaos}.

In this paper we use the second physical interpretation 
of the coordinate fractional integration.
This interpretation follows from the fractional measure
of space \cite{chaos} that is used in the fractional integrals.
We consider the mass fractional dimension
and the fractional generalizations of the equation 
that defines the medium mass.
We use the fractional integrals to describe fractal media 
with noninteger mass dimensions \cite{PLA05}.
Using fractional integrals, we derive the fractional generalization 
of Chapman-Kolmogorov equation (Smolukhovski equation).  
In this paper the fractional generalization of Fokker-Planck equation 
is derived from the fractional Chapman-Kolmogorov equation. 
Using the Fourier transform, we derive the Fokker-Planck-Zaslavsky 
equations \cite{Zas2,Zas3} that have fractional coordinate derivatives.
The Fokker-Planck equation for the fractal media is an equation 
with fractional derivatives in the dual space. 

In Sec. 2, the definition of the mass dimension 
and the fractional generalization of the media mass 
equation are considered. In Sec. 3, we derive
the fractional generalization of the average values equation.
In Sec. 4, the fractional Chapman-Kolmogorov equation
is derived by using fractional integration. 
In Sec. 5, the fractional Fokker-Planck equation for the fractal media 
is derived from the suggested fractional Chapman-Kolmogorov equation. 
In Sec. 5, we use only the fractional power coordinate derivatives. 
The fractional derivatives are used only in Sec. 7.
In Sec.6, the stationary solution of the Fokker-Planck equation
for fractal media are derived. 
In Sec. 7, we consider Fokker-Planck-Zaslavsky  equations with
fractional coordinate derivatives. We derived these equations 
by the Fourier transform of the Fokker-Planck equations for fractal media.

\section{Mass Fractal Media and Fractional Integral}

Equations that define the fractal dimensions have 
the passage to the limit. This passage makes difficult 
the practical application to the real fractal media.
The other dimensions, which can be easy calculated from
the experimental data, are used in empirical investigations.
For example, the mass fractal dimension \cite{Mand,Schr} 
can be easy calculated.
The properties of the fractal media like mass obeys a power law relation
\be \label{MR} M(R) =kR^{D_m} , \ee
where $M$ is the mass of fractal medium, $R$ is a box size (or a sphere radius),
and $D_m$ is a mass fractal dimension. 
The amount of mass of a medium inside a box of size $R$
has a power law relation (\ref{MR}).

Fractal dimension can be best calculated by box counting 
which means drawing a box of size R and counting the mass inside. 
To calculate the mass fractal dimension, take the logarithm of 
both sides of Eq. (\ref{MR}): $D_m=ln(M)/ln(R)$.
Log-log plot of $M$ and $R$ gives us the slope $D_m$, 
the fractal dimension. 
When we graph ln(M) as a function of ln(R) we get a value 
of about $D_m$ which is the fractal dimension of fractal media.

Let us prove that power law relation (\ref{MR}) can be naturally 
derived by using the fractional integral approach \cite{chaos}.
Let us derive the fractional generalization of the equation,
which defines the mass of medium that is distributed on 
the line $-(R/2)<x<(R/2)$ and has the form
\be M_1(R)= \int^{R/2}_{-R/2} \rho(x) dx , \ee
where $\rho(x) \in L_1(R^1)$.
This equation can be written in the following form:
\be \label{If}
M_1(R)=\int^y_{-R/2} \rho(x) dx +\int^{R/2}_y \rho(x) dx . \ee
Using the notations \cite{SKM} for fractional integrals
\[ (I^{\alpha}_{R/2+}\rho )(y)=
\frac{1}{\Gamma (\alpha)} \int^{y}_{-R/2}
\frac{\rho(x)dx}{(y-x)^{1-\alpha}} , \]
\[ \label{I-0} (I^{\alpha}_{R/2-}\rho)(y)=
\frac{1}{\Gamma (\alpha)} \int^{R/2}_{y}
\frac{\rho(x)dx}{(x-y)^{1-\alpha}} , \]
where $-(R/2) <y<(R/2)$,
we can rewrite Eq. (\ref{If}) in the form
\[ M_1(R)=(I^{1}_{R/2+}\rho)(y)+(I^{1}_{R/2-}\rho)(y) . \]
The fractional generalization of
this equation is defined by
\be \label{M1} M_{\alpha}(R)=
(I^{\alpha}_{R/2+}\rho)(y)+(I^{\alpha}_{R/2-}\rho)(y) , \ee
where $\rho(x) \in L_p(R^1)$, and $1<p<1/\alpha$.
The initial points in Eq. (\ref{M1}) are set to zero ($y=0$).
As the result,we can rewrite Eq. (\ref{M1}) in an equivalent form
\[ M_{\alpha}(R) =\frac{1}{\Gamma(\alpha)} \int^{R/2}_0
[\rho(x)+\rho(-x)] x^{\alpha-1} dx . \] 

In order to describe the fractal media,  
we suggest to use the fractional continuous medium model \cite{PLA05}. 
We proposes \cite{PLA05} to describe fractal media by a "fractional" continuous
medium model where all characteristics are defined everywhere in the volume 
but they follow some generalized equations which are derived by 
using fractional integrals.
  
If we use Eq. (\ref{M1}) and 
use the constant density distribution ($\rho(x)=\rho_0=const$)
for homogeneous fractal media, then we get
\[ M_{\alpha}(R)=\frac{\rho_0 2^{1-\alpha}}{\alpha\Gamma(\alpha)} R^{\alpha} .\]
We have that the mass fractal dimension $D_m=\alpha$, 
and $k=\rho_0 2^{1-\alpha}/[\alpha \Gamma(\alpha)]$ in Eq. (\ref{MR}). 
The fractional generalization of the equations for fractal media
in the three-dimensional space are considered in Ref. \cite{PLA05}
and in the Appendix of this paper.

The fractional integrals can be used to describe fractal
media with noninteger mass dimensions \cite{PLA05}. 
The fractional integrals can be used not only to calculate 
the mass dimensions of fractal media. Fractional integration can be used
to describe the dynamical processes in the fractal media.
Using the fractional integrals \cite{chaos},  
we can derive the fractional generalizations of the 
Chapman-Kolmogorov and Fokker-Planck equations.

\section{Fractional Average Values}

In order to derive the fractional analog of average value
we consider the fractional integral for the function $A(x)$.
The usual average value
\be <A>_1= \int^{+\infty}_{-\infty} A(x)\rho(x) dx  \ee
can be written in the form
\be \label{If2}
<A>_1=\int^y_{-\infty} A(x)\rho(x) dx +\int^{+\infty}_y A(x)\rho(x) dx . \ee
Using the following notations for the fractional integrals,
\be  \label{I+} (I^{\alpha}_{+}f)(y)=
\frac{1}{\Gamma (\alpha)} \int^{y}_{-\infty}
\frac{f(x)dx}{(y-x)^{1-\alpha}} , \ee
\be \label{I-} (I^{\alpha}_{-}f)(y)=
\frac{1}{\Gamma (\alpha)} \int^{+\infty}_{y}
\frac{f(x)dx}{(x-y)^{1-\alpha}} , \ee
the average value (\ref{If2}) can be rewritten in the form
\[ <A>_1(y)=(I^{1}_{+}A\rho)(y)+(I^{1}_{-}A\rho)(y) . \]
The fractional generalization of this equation can 
be defined by the equation
\be \label{Aa} <A>_{\alpha}(y)=
(I^{\alpha}_{+}A\rho)(y)+(I^{\alpha}_{-}A\rho)(y) . \ee
Note that Eq. (\ref{Aa}) can be written \cite{chaos} in the 
equivalent form, 
\be \label{FI2} <A>_{\alpha}(y)=
\int^{+\infty}_{0} [ (A\rho)(y-x)+ (A\rho)(y+x) ] d\mu_{\alpha}(x) , \ee
where we use
\[ d\mu_{\alpha}(x)=\frac{|x|^{\alpha-1} dx}{\Gamma(\alpha)}=
\frac{d x^{\alpha}}{\alpha \Gamma(\alpha)} .\]
Here we use the following notation for fractional power of
coordinates:
\be \label{xa}
x^{\alpha} =\beta(x) (x)^{\alpha}= sgn(x) |x|^{\alpha} , \ee
where $\beta(x)=[sgn(x)]^{\alpha-1}$. 
The function $sgn(x)$ is equal to $+1$ for $x\ge0$,
and this function is equal to $-1$ for $x<0$.

In order to have the symmetric limits of the integral we consider
average value (\ref{FI2}) in the form
\be \label{FI5} <A>_{\alpha}(y)= \frac{1}{2}
\int^{+\infty}_{-\infty} [ (A\rho)(y-x)+ (A\rho)(y+x)] d\mu_{\alpha}(x) . \ee
If $\alpha=1$, then we have the usual equation for the average value. 

Let us introduce some notations in order to simplify 
the equation for the fractional average value. 
We use the integral operators $\hat I^{\alpha}_{x}$
that is defined by
\be \label{hatI} \hat {\bf I}^{\alpha}_{x} f(x)=\frac{1}{2}
\int^{+\infty}_{-\infty}  \Bigl[ f(x)+f(-x) \Bigr] d \mu_{\alpha} (x) .\ee
In this case, fractional generalization of the average 
value (\ref{FI5}) can be rewritten in the form
\be \label{15} 
<A>_{\alpha}=\hat {\bf I}^{\alpha}_{x}  \ A(x)\rho(x) . \ee
We will use the initial points in Eq. (\ref{M1}) are set to zero ($y=0$).
Note that the fractional normalization condition \cite{chaos}
is a special case of this definition of average values $<1>_{\alpha}=1$.

\section{Fractional Chapman-Kolmogorov Equation}

The Fokker-Planck equation
can be derived from so-called Chapman-Kolmogorov equation
(also known as Smolukhovski equation) \cite{Kolm1,Kolm2,Levy}.
This equation may be interpreted as the condition of
consistency of distribution functions of different orders.

Kolmogorov \cite{Kolm1,Kolm2} derived a kinetic equation using a 
special scheme and conditions that are important for kinetics. 
Let $W(x,t;x_0,t_0)$ be a probability density of having a particle 
at the position $x$ at time $t$ if the particle was at the position $x_0$ 
at time $t_0\le t$. 

Denote by $\rho(x,t)$ the distribution functions for the given time $t$. 
Let us consider two well-known identities
\be  \label{rpr0} \rho(x,t) = \int^{+\infty}_{-\infty} d{x'}
\ W(x,t|x',t') \rho(x',t') , 
\quad \int^{+\infty}_{-\infty} \rho(x,t)=1 . \ee
Using notation (\ref{hatI}), we can rewrite 
these equations in the form
\[  \rho(x,t) =\hat {\bf I}^{1}_{x'}  \ W(x,t|x',t') \rho(x',t') , \quad
\hat {\bf I}^{1}_x  \ \rho(x,t)=1 . \]
Therefore we can get
the fractional generalization of Eqs. (\ref{rpr0}) in the form
\be  \label{rpr} \rho(x,t) =
\hat {\bf I}^{\alpha}_{x'} \ W(x,t|x',t') \rho(x',t') . \ee
This equation is
the definition of conditional distribution function $W(x,t|x',t')$ for 
the fractal medium referring to different time instants.
The normalization conditions for  the functions $W(x,t|x',t')$ 
and $\rho(x,t)$ have the form 
\be \label{nc-p} \hat {\bf I}^{\alpha}_x  \ W(x,t|x',t')=1 , \quad
\hat {\bf I}^{\alpha}_x  \ \rho(x,t)=1 . \ee
Substituting into the right-hand side of Eq. (\ref{rpr})
the value of $\rho(x',t')$
expressed via the distribution $\rho(x_0,t_0)$ at an earlier time,
\be  \label{rpr-2} \rho(x',t') =
\hat {\bf I}^{\alpha}_{x_0}  \ W(x',t'|x_0,t_0) \rho(x_0,t_0) , \ee
we obtain the integral relation which includes the
intermediate point $x'$,
\be \label{20} \rho(x,t)=
\hat {\bf I}^{\alpha}_{x'} \ \hat {\bf I}^{\alpha}_{x_0}
 \  W(x,t|x',t')W(x',t'|x_0,t_0) \rho(x_0,t_0) . \ee
Using Eq. (\ref{20}), and Eq. (\ref{rpr}) in the form
\be  \label{rpr-3} \rho(x,t) =
\hat {\bf I}^{\alpha}_{x_0}  \ W(x,t|x_0,t_0) \rho(x_0,t_0) , \ee
we can derive a closed equation for transition
probabilities
\[ \hat {\bf I}^{\alpha}_{x_0}  \ W(x,t|x_0,t_0) \rho(x_0,t_0) =
\hat {\bf I}^{\alpha}_{x'}  \ \hat {\bf I}^{\alpha}_{x_0}  \
W(x,t|x',t')W(x',t'|x_0,t_0) \rho(x_0,t_0) . \]
Since the equation holds for arbitrary $\rho(x_0,t_0)$,
we may equate the integrand. As the result, we get
the fractional Chapman-Kolmogorov equation 
\be W(x,t|x_0,t_0)=\hat {\bf I}^{\alpha}_{x'}  \
W(x,t|x',t')W(x',t'|x_0,t_0) . \ee
This equation describes the Markov-type process in the fractal medium 
that is described by the continuous medium model \cite{PLA05}. 
Using this fractional equation we can derive the 
fractional Fokker-Plank equation.

\section{Fractional Fokker-Planck Equation from \\
Fractional Chapman-Kolmogorov Equation}

Let us consider the fractional generalization average value (\ref{15})
for the function $A(x)$, 
\be \label{av1} <A>_{\alpha}=\hat {\bf I}^{\alpha}_x  \ A(x) \rho(x,t) . \ee
We will use the initial points are set to zero ($y=0$).
Using Eq. (\ref{rpr}) in the form
\be  \label{rpr2} \rho(x,t) =
\hat {\bf I}^{\alpha}_{x_0}  \ W(x,t|x_0,t_0) \rho(x_0,t_0) , \ee
we get the average value
\[ <A>_{\alpha}=\hat {\bf I}^{\alpha}_x  \ A(x)\ 
\hat {\bf I}^{\alpha}_{x_0}  \ W(x,t|x_0,t_0) \rho(x_0,t_0) . \]
We can rewrite this equation in the form
\be \label{av3} <A>_{\alpha}=\hat {\bf I}^{\alpha}_{x_0}  \
\rho(x_0,t_0) \ \hat {\bf I}^{\alpha}_x  \ A(x) \ W(x,t|x_0,t_0)  . \ee

Let us consider the Taylor expansion of function $A(x)$.
If we use this expansion in the form
\be \label{TE1}  A(x)=A(x_0+\Delta x)=A(x_0)+
\Bigl(\frac{\partial A(x)}{\partial x}\Bigr)_{x_0} \Delta x+ 
\frac{1}{2}\Bigl(
\frac{\partial^2 A(x)}{(\partial x)^{2}}\Bigr)_{x_0}
(\Delta x)^2+..., \ee
where $\Delta x=x-x_0$, then the fractional integration by
parts of Eq. (\ref{av3}) cannot be realized in the simple form.
It can be assumed \cite{chaos} that the function $A$ has the form
$A=A(x^{\alpha})$. The Taylor expansion can be used in the form
\be \label{TE2}
A(x^{\alpha})=A(x^{\alpha}_0+\Delta x^{\alpha})=A(x^{\alpha}_0)+
\Bigl(\frac{\partial A(x^{\alpha})}{\partial x^{\alpha}}\Bigr)_{x_0}
\Delta x^{\alpha}+ 
\frac{1}{2}\Bigl(\frac{\partial^2 A(x^{\alpha})}{(\partial x^{\alpha})^2}
\Bigr)_{x_0} (\Delta x^{\alpha})^2+..., \ee
where $\Delta x^{\alpha}=x^{\alpha}-x^{\alpha}_0$, and 
$x^{\alpha}=sgn(x)|x|^{\alpha}$ is defined by Eq. (\ref{xa}).
Here and later, we use the following notations for derivatives of 
fractional power of coordinates:
\be \label{am}
\frac{\partial}{\partial x^{\alpha}}= \frac{|x|^{1-\alpha}}{\alpha}
\frac{\partial}{\partial x} .
\ee
Note that these derivatives are the derivatives of fractal 
order in the dual space ("momentum representation").  
The expansion (\ref{TE2}) is equivalent to the usual Taylor expansion.
In order to prove this proposition, we consider the Taylor expansion
for $x^{\alpha}$ that has the form
\be \label{Deltax} \Delta x^{\alpha}=x^{\alpha}-x^{\alpha}_0=
\Bigl(\frac{\partial x^{\alpha}}{\partial x} \Bigr)_{x_0} 
\Delta x+\frac{1}{2} 
\Bigl(\frac{\partial^2 x^{\alpha}}{(\partial x)^2} 
\Bigr)_{x_0} (\Delta x)^2 +... \ . \ee
Substituting Eq. (\ref{Deltax}) into Eq. (\ref{TE2}), 
we get the usual Taylor expansion (\ref{TE1}). 
If we use the usual Taylor expansion (\ref{TE1}), then the integration by parts
in Eq. (\ref{av3}) is more complicated. If we use the Taylor expansion for the 
fractional power of coordinates, then the integration by parts 
in Eq. (\ref{av3}) can be realized in the simple form, 
\[ \hat {\bf I}^{\alpha}_x B(x) 
\frac{\partial A(x^{\alpha})}{\partial x^{\alpha}}=
\int^{+\infty}_{-\infty} \frac{dx^{\alpha}}{\alpha \Gamma(\alpha)} B(x)
\frac{\partial A(x^{\alpha})}{\partial x^{\alpha}}=\]
\[ =\Bigl( B(x)A(x) \Bigr)^{+\infty}_{-\infty}-
\int^{+\infty}_{-\infty} 
\frac{dx^{\alpha}}{\alpha \Gamma(\alpha)} A(x^{\alpha})
\frac{\partial B(x)}{\partial x^{\alpha}} . \]

Substituting Eq. (\ref{TE2}) in Eq. (\ref{av3}), we get
\[ <A>_{\alpha}= \hat {\bf I}^{\alpha}_{x_0}  \ A(x^{\alpha}_0) \rho(x_0,t_0)
\hat {\bf I}^{\alpha}_x  \  W(x,t|x_0,t_0) +  \]
\[ +\hat {\bf I}^{\alpha}_{x_0}  \
\Bigl(\frac{\partial A(x^{\alpha})}{\partial x^{\alpha}}\Bigr)_{x_0} 
\rho(x_0,t_0) \hat {\bf I}^{\alpha}_x  \ \Delta x^{\alpha}  W(x,t|x_0,t_0)+\]
\be \label{av4}
+\frac{1}{2} \hat {\bf I}^{\alpha}_{x_0}  \
\Bigl(\frac{\partial^2 A(x^{\alpha})}{(\partial x^{\alpha})^2}\Bigr)_{x_0}
\rho(x_0,t_0)  
\hat {\bf I}^{\alpha}_x  \ (\Delta x^{\alpha})^2  W(x,t|x_0,t_0)+...  \ee
Let us introduce the following functions:
\be \label{Pn} P_n(x_0,t,t_0)=\hat {\bf I}^{\alpha}_x  \ (\Delta x^{\alpha})^n
W(x,t|x_0,t_0) . \ee
Using (\ref{Pn}) and (\ref{nc-p}), we can rewrite Eq. (\ref{av4}) in the form
\[ <A>_{\alpha}= \hat {\bf I}^{\alpha}_{x_0} 
\ A(x^{\alpha}_0) \rho(x_0,t_0) +  
\hat {\bf I}^{\alpha}_{x_0}  \
\Bigl(\frac{\partial A(x^{\alpha})}{\partial x^{\alpha}}\Bigr)_{x_0} 
\rho(x_0,t_0) P_1(x_0,t,t_0)+\]
\be \label{av5} +\frac{1}{2} \hat {\bf I}^{\alpha}_{x_0}  \
\Bigl(\frac{\partial^2 A(x^{\alpha})}{(\partial x^{\alpha})^2}\Bigr)_{x_0}
\rho(x_0,t_0) P_2(x_0,t,t_0)+... \ .
\ee
Using (\ref{av1}) in the form
\[ <A>_{\alpha}=
\hat {\bf I}^{\alpha}_{x_0}  \ A(x^{\alpha}_0) \rho(x_0,t) , \]
for the left hand side of Eq. (\ref{av5}), we can rewrite this equation 
\[ \hat {\bf I}^{\alpha}_{x_0}  \
A(x_0)\Bigl(\rho(x_0,t)-\rho(x_0,t_0)\Bigr)=  \]
\be \label{av7} 
=\hat {\bf I}^{\alpha}_{x_0}  \
\Bigl(\frac{\partial A(x)}{\partial x^{\alpha}}\Bigr)_{x_0} \rho(x_0,t_0)
P_1(x_0,t,t_0)+ \frac{1}{2} \hat {\bf I}^{\alpha}_{x_0}  \
\Bigl(\frac{\partial^2 A(x)}{(\partial x^{\alpha})^2}\Bigr)_{x_0}
\rho(x_0,t_0) P_2(x_0,t,t_0)+... \ . \ee
The final step is an assumption that
is usually called Kolmogorov condition \cite{Kolm1,Kolm2}. 
We may assume that the following finite limits exist:
\[ \lim_{\Delta t \rightarrow 0} \frac{P_1(x,t,t_0)}{\Delta t}= a(x,t_0) , \]
\[ \lim_{\Delta t \rightarrow 0} \frac{P_2(x,t,t_0)}{\Delta t}= b(x,t_0) , \]
and
\[ \lim_{\Delta t \rightarrow 0} \frac{P_n(x,t,t_0)}{\Delta t}= 0 , \]
where $n \ge 3$.
It is due to the Kolmogorov conditions that irreversibility
appears at the final equation.
Multiplying both sides of Eq. (\ref{av7}) by $1/\Delta t$ and
consider the limit, we obtain
\[ \hat {\bf I}^{\alpha}_{x_0}  \ A(x^{\alpha}_0)
\Bigl(\frac{\partial \rho(x_0,t)}{\partial t} \Bigr)_{t_0} = \] 
\[ =\hat {\bf I}^{\alpha}_{x_0}  \
\Bigl(\frac{\partial A(x^{\alpha})}{\partial x^{\alpha}}\Bigr)_{x_0} 
\rho(x_0,t_0) a(x_0,t_0) 
+\frac{1}{2} \hat {\bf I}^{\alpha}_{x_0}  \
\Bigl(\frac{\partial^2 A(x^{\alpha})}{(\partial x^{\alpha})^2}\Bigr)_{x_0}
\rho(x_0,t_0) b(x_0,t_0) . \]
Integrating by parts, we obtain
\be \label{ip1}
\hat {\bf I}^{\alpha}_{x}  \
\frac{\partial A(x^{\alpha})}{\partial x^{\alpha}} \rho(x,t) a(x,t) 
=- \hat {\bf I}^{\alpha}_{x}  \ A(x^{\alpha})
\frac{\partial (\rho(x,t) a(x,t))}{\partial x^{\alpha}} , \ee

\be \label{ip2}
\hat {\bf I}^{\alpha}_{x}  \
\frac{\partial^2 A(x^{\alpha})}{(\partial x^{\alpha})^2}
\rho(x,t) b(x,t)
=\hat {\bf I}^{\alpha}_{x}  \ A(x^{\alpha})
\frac{ \partial^2 (\rho(x,t) b(x,t)) }{ (\partial x^{\alpha})^2 } . \ee
Here we use
\[ \lim_{x\rightarrow \pm \infty} \rho(x,t)=0 .\]
Then we have the equation
\[ \hat {\bf I}^{\alpha}_{x}  \ A(x^{\alpha})
\Bigl( \frac{\partial \rho(x,t)}{\partial t}+
\frac{\partial (\rho(x,t)a(x))}{\partial x^{\alpha}}-\frac{1}{2}
\frac{\partial^2 (\rho(x,t)b (x)) }{ (\partial x^{\alpha})^2 }
\Bigr)=0. \]
Since the function $A=A(x^{\alpha})$ is an arbitrary function, then 
we have the following equation
\be \label{FP} \frac{\partial \rho(x,t)}{\partial t}+
\frac{\partial (\rho(x,t)a(x,t))}{\partial x^{\alpha}} -\frac{1}{2}
\frac{\partial^2 (\rho(x,t)b (x,t))}{ (\partial x^{\alpha})^2 }=0  , \ee
that is the Fokker-Planck equation for fractal media. 
This equation is derived from the fractional generalization of the
average value and fractional normalization condition, which use 
the fractional integrals. 
The Fourier transform of this equation has the fractional derivatives.
We consider the equations with fractional derivative in Section 7.
Therefore this equation can be called the fractional 
Fokker-Planck equation. The Fokker-Planck equation with 
fractional derivatives we call Fokker-Planck-Zaslavsky equation, since
this equation was suggested by Zaslavsky \cite{Zas,Zas2,Zas3}.

\section{Stationary Solutions}

In this section, we consider simplest solutions of the Fokker-Planck equations
for the fractal media.
Let us consider the stationary Fokker-Planck equation for fractal media
\be \label{sFP1} \frac{\partial (\rho(x,t)a(x,t))}{\partial x^{\alpha}} -\frac{1}{2}
\frac{\partial^2 (\rho(x,t)b (x,t))}{(\partial x^{\alpha})^2}=0  . \ee
This equation can be rewritten in the form
\be \label{sFP2} 
\frac{\partial}{\partial x^{\alpha}} \Bigl( \rho(x,t)a(x,t)-\frac{1}{2}
\frac{\partial (\rho(x,t)b (x,t))}{\partial x^{\alpha}}\Bigr)=0  . \ee
Obviously, we get the relation 
\be \label{sFP3} 
\rho(x,t)a(x,t) -\frac{1}{2}
\frac{\partial (\rho(x,t)b (x,t))}{\partial x^{\alpha}}=const  . \ee
Supposing that the constant is equal to zero, we get
\be \label{sFP4} 
\frac{\partial (\rho(x,t)b(x,t))}{\partial x^{\alpha}}=
\frac{2a(x,t)}{b(x,t)} (\rho(x,t)b(x,t)) , \ee
or, in an equivalent form
\be \label{sFP5} 
\frac{\partial \ ln (\rho(x,t)b(x,t))}{\partial x^{\alpha}}=
\frac{2a(x,t)}{b(x,t)} . \ee
The solution of this equation is
\be \label{sFP6} 
 ln (\rho(x,t)b(x,t)) = \int \frac{2a(x,t)}{b(x,t)} dx^{\alpha} +const. \ee
As the result, we have the following solution of the
stationary Fokker-Planck equation for fractal media
\be \label{sFP7} 
\rho(x,t)= \frac{N}{b(x,t)} exp \ 2 \int \frac{a(x,t)}{b(x,t)} dx^{\alpha} , \ee
where the coefficient $N$ is defined by the normalization condition.

Let us consider the special cases of this solution. \\
(1) If $a(x)=k$ and $b(x)=-D$, then Fokker-Planck equation 
for fractal media has the form
\be \label{FP-case1} \frac{\partial \rho(x,t)}{\partial t}+
k\frac{\partial \rho(x,t)}{\partial x^{\alpha}} +\frac{D}{2}
\frac{\partial^2 \rho(x,t)}{(\partial x^{\alpha})^2}=0  . \ee
The stationary solution of this equation is the 
following distribution function:
\be \label{case1} \rho(x,t)= N_1 
exp \left( - \frac{2 k |x|^{\alpha}}{D} \right) . \ee
(2) If $a(x)=k|x|^{\beta}$ and $b(x)=-D$, then Fokker-Planck equation 
for fractal media has the form
\be \label{FP-case2} \frac{\partial \rho(x,t)}{\partial t}+
k\frac{\partial  |x|^{\beta}\rho(x,t)}{\partial x^{\alpha}} +\frac{D}{2}
\frac{\partial^2 \rho(x,t)}{(\partial x^{\alpha})^2}=0  . \ee
The stationary solution of this equation is the 
following distribution function:
\be \label{case2} \rho(x,t)= N_2 
exp \left( - \frac{2 \alpha k |x|^{\alpha+\beta}}{(\alpha+\beta)D} \right) . 
\ee
If $\beta=\alpha$, we get
\be \label{case2b} \rho(x,t)= N_2 
exp \left( - \frac{k}{D} |x|^{2\alpha} \right) . \ee
If $\alpha+\beta=2$, we have
\be \label{case2c} \rho(x,t)= 
N_2 exp \left( - \frac{\alpha k}{D} x^{2} \right) . \ee
(3) If $a(x)=k|x|^{\beta}$ and $b(x)=-D|x|^{-\gamma}$, then 
the stationary solution of the Fokker-Planck equation for fractal 
media is the following distribution function:
\be \label{case3} 
\rho(x,t)= N_3 |x|^{\gamma} 
exp \left( - \frac{2 \alpha k}{(\alpha+\beta+\gamma)D} 
|x|^{\alpha+\beta+\gamma}  \right). 
\ee
(4) If the coefficient $a(x)$ is defined by
\[ a(x)=\frac{\partial U(x)}{\partial x^{\alpha}}=\frac{|x|^{1-\alpha}}{\alpha} 
\frac{\partial U(x)}{\partial x},  \] 
and $b=-D$, then we get the following stationary solution:
\[ \rho(x,t)= N_4 exp \left( - \frac{U(x)}{D} \right). \]

Let us consider the Fokker-Planck equation for fractal media (\ref{FP})
with $a(x)=k|x|^{\alpha}$ and $b=-D$. 
The stationary solution (\ref{case2b}) has the form
\[\rho(x)=\Bigl(\frac{k}{\pi D}\Bigr)^{1/2} e^{-k x^{2\alpha}/ D} . \]
The general solution can be represented in the form
\[ \rho(x,t)=\sum^{+\infty}_{n=0}
\sqrt{\frac{k}{2^n n!\pi D}} e^{-k x^{2\alpha}/ D} 
H_n(x^{\alpha \sqrt{k/ D}}) e^{-nkt} A_n , \]
where
\[ A_n=\sqrt{\frac{1}{2^n n!}}\  \hat {\bf I}^{\alpha}_x \ p(x,0) 
H_n(x^{\alpha}\sqrt{k/ D}) . \]
Note that 
\[ <x^{\alpha}(t)x^{\alpha}(0)>_{\alpha}=\frac{D}{2k} e^{-kt} . \]


\section{Fokker-Planck-Zaslavsky Equations}

The fractional generalization of the Fokker-Planck equations
was suggested by Zaslavsky in Ref. \cite{Zas}.
These equations have fractional derivatives and 
can be used to describe the dynamical processes 
on fractals \cite{Zas2,Zas3,Zas4}. 
The fractals can be realized in nature as fractal media. 
Therefore it is interesting to derive Fokker-Planck-Zaslavsky  
equation as an equation for fractal media \cite{PLA05}. 

We derive the Fokker-Planck-Zaslavsky equations that have coordinate
fractional derivatives in the dual space. 
The Fokker-Planck equations for fractal media are equations 
with fractional derivatives in the dual space ("momentum representation").
Let us use Eq. (\ref{FP}) and  the Fourier transform ${\cal F}$ 
of the coordinates that leads us to the Reisz derivatives.  

Using Eq. (\ref{am}), we can rewrite the Fokker-Planck equation 
for fractal media (\ref{FP}) in the form with usual coordinate derivatives 
$\partial_x=\partial / \partial x$:
\[ \partial_t \rho(t,x)+\frac{|x|^{1-\alpha}}{\alpha}
\partial_x (a(x)\rho(t,x)) -
\frac{(1-\alpha)|x|^{1-2\alpha}}{2\alpha^2}
\partial_x (b(x)\rho(t,x))-
\frac{|x|^{2-2\alpha}}{2\alpha^2}
\partial^2_x (b(x)\rho(t,x))=0 . \]
Let us consider two special cases, (1) the coefficients $a(x)$ and $b(x)$ 
are constant, (2) the coefficients $a(x)$ and $b(x)$ are connected
by the equation $a(x)=(1/2) \partial b(x) / \partial x^{\alpha}$. \\

(1) If the functions $a(x)$ and $b(x)$ do not have the coordinate
dependence ($a=k$ and $b=-D$), then we get the equation
\be \label{FP-case1-1} \frac{\partial \rho(x,t)}{\partial t}+
k\frac{\partial \rho(x,t)}{\partial x^{\alpha}} +\frac{D}{2}
\frac{\partial^2 \rho(x,t)}{(\partial x^{\alpha})^2}=0  . \ee
This equation can be rewritten in the form
\[ \partial_t \rho(t,x)+\frac{k|x|^{1-\alpha}}{\alpha}
\partial_x \rho(t,x) -
\frac{D(1-\alpha)|x|^{1-2\alpha}}{2\alpha^2} \partial_x \rho(t,x)-
\frac{D|x|^{2-2\alpha}}{2\alpha^2}\partial^2_x \rho(t,x)=0 . \]

The Fourier transform ${\cal F}$ of the coordinates $|x|^{\beta}$ leads us 
to the Reisz derivatives \cite{SKM}
\[ {\bf D}^{\beta}_y=\frac{1}{2 cos(\pi \beta/2)}\Bigl(
\frac{1}{\Gamma(1-\beta)} \frac{d}{dy} \int^y_{-\infty}
\frac{f(z)dz}{(y-z)^{\beta}}+
\frac{-1}{\Gamma(1-\beta)} \frac{d}{dy} \int^{+\infty}_y
\frac{f(z)dz}{(z-y)^{\beta}} \Bigr) \]
by the equations
\[ {\cal F}[|x|^{\beta} \rho(t,x)]=-{\bf D}^{\beta}_y
\tilde \rho(t,y) , \]
where $\tilde \rho(t,y)={\cal F}[\rho(t,x)]$. 
As the result, we have the fractional Fokker-Planck equation
with fractional derivatives in the dual space
\be \label{53} \partial_t \tilde \rho(t,y)+\frac{k}{\alpha}
{\bf D}^{1-\alpha}_y \Bigl(iy \tilde \rho(t,y)\Bigr) +
\frac{D(1-\alpha)}{2\alpha^2} {\bf D}^{1-2\alpha}_y
\Bigl(iy \tilde \rho(t,y)\Bigr)+\frac{D}{2\alpha^2} {\bf D}^{2-2\alpha}_y
\Bigl(y^2 \tilde \rho(t,y) \Bigr)=0  \ee
for the function $\tilde \rho(t,y)$. 
Here we use ${\cal F}[\partial^n_x \rho(t,x)]= (-iy) \tilde \rho(t,y)$. 
Equation (\ref{53}) is a dual form of Fokker-Planck 
equation (\ref{FP-case1-1}).
As the result, we have fractional Fokker-Planck in the dual space. \\

(2) After using the relation
\be a(x)=\frac{1}{2}\frac{\partial b(x)}{\partial x^{\alpha}} , \ee
we get the diffusion equation (\ref{FP}) in the form 
\[ \frac{\partial \rho(x,t)}{\partial t}-\frac{1}{2}
\frac{\partial}{\partial x^{\alpha}} \Bigl( D(x)
\frac{\partial \rho(x,t)}{\partial x^{\alpha}}\Bigr)=0  \]
with diffusion coefficient $D(x)=b(x)$. 
Here derivatives $\partial / \partial x^{\alpha}$ are 
defined by Eq. (\ref{am}). 
Using Eq. (\ref{am}), we have
\[ \frac{\partial \rho(x,t)}{\partial t}-\frac{|x|^{1-\alpha}}{2\alpha^2}
\partial_x \Bigl( |x|^{1-\alpha} D(x)
\partial_x \rho(x,t) \Bigr)=0 . \]
If the diffusion coefficient $D(x)$ is described by the power law
\[ |x|^{1-\alpha} D(x)=\gamma |x|^\beta , \]
then we have the diffusion equation in the form
\[ \partial_t \rho(x,t)-\frac{\gamma}{2\alpha^2}|x|^{1-\alpha}
\partial_x \Bigl( |x|^{\beta} \partial_x  \rho(x,t) \Bigr)=0 . \]
The Fourier transform ${\cal F}$ of the coordinates leads us 
the following fractional Fokker-Planck equation
with the Reisz fractional derivatives 
\[ \partial_t \tilde \rho(y,t)-\frac{\gamma}{2\alpha^2} {\bf D}^{1-\alpha}_y
\Bigl( y {\bf D}^{\beta}_y y  \tilde \rho(y,t) \Bigr)=0 . \]
If $\beta=0$, then we have 
the fractional kinetic equation
\be \partial_t \tilde \rho(y,t)= {\bf D}^{1-\alpha}_y ( {\cal A}(y) \tilde \rho(y,t) ) , \ee
where ${\cal A}(y)=(\gamma / 2\alpha^2) y^2$. 
As the result, we get the equations with fractional derivatives 
that have Fokker-Planck-Zaslavsky form \cite{Zas2,Zas3} in the dual space.

\newpage
\section{Conclusion}

The application of fractals in physics is far ranging, from the 
dynamics of fluids in porous media to resistivity 
networks in electronics. 
The cornerstone of fractals is the meaning of dimension, 
specifically the fractal dimension. 
Fractal dimension can be best calculated by box counting 
method which means drawing a box of size $R$ 
and counting the mass inside. 
When we graph ln($M$) as a function of ln($R$) we get a value 
of about $D_m$ which is the fractal dimension of fractal medium. 
The mass fractal dimensions of the media 
can be easy measured by experiments. 
The experimental determination of the mass fractal dimension 
can be realized by the usual  box counting methods.
The mass dimensions were measured for porous materials
\cite{Por1,Por2}, polymers \cite{P}, and colloid agregates \cite{CA}.
Fractal models of porous media are enjoying considerable
popularity \cite{AT,PMRM,BD,PBR,RC}. 
This is due in part to the relatively 
small number of parameters that can define a fractal porous medium 
of great complexity and rich structure.

We suppose that the concept of fractional integration
provides an alternative approach to describe the fractal media. 
In this paper we prove that fractal media can be described
by using the fractional integrals. 
We consider the fractional generalizations of 
the equation that defines the medium mass.
This fractional generalization allows us to realize the 
consistent description of the fractal mass dimension of the media \cite{PLA05}. 
The fractional integrals can be used in order to describe 
the dynamical processes in the fractal media. 
Fractional integration approach is potentially more useful 
for physics of fractal media than traditional methods that 
use the integer integration.
In this paper fractional generalizations of the
average value is derived.
Using fractional integrals, we derive the fractional generalization of the 
Chapman-Kolmogorov equation.
In this paper we derive the Fokker-Planck equations 
for the fractal media from the suggested 
fractional Chapman-Kolmogorov equation.
Using the Fourier transform, we get the Fokker-Planck-Zaslavsky equations 
that have coordinate fractional derivatives in the dual space 
("momentum representation"). 

Note that the Liouville equation is a cornerstone of 
the statistical mechanics \cite{Is,RL,F}.  
The fractional generalization of the Liouville equation 
was suggested in Ref. \cite{chaos}.
The fractional generalization of Liouville equation 
allows us to derive the fractional generalization of the Bogoliubov equations,
Fokker-Planck equation, Enskog transport equation, and hydrodynamic equations. 
Using fractional analog of the Liouville equation \cite{chaos}, 
we can derive the fractional Fokker-Planck equation for the phase space. 

Using the methods suggested in Refs. \cite{Tarpla1,Tarmsu,Tarsam},
we can realize the Weyl quantization of the suggested 
fractional equations. \\

\section*{Acknowledgment}

The author would like to thank 
Professor G.M.  Zaslavsky for very useful discussions.

\section*{Appendix}

In many problems the real fractal structure of matter 
can be disregarded and the medium can be replaced by  
some "fractional" continuous mathematical model \cite{PLA05}. 
In order to describe the media with 
noninteger mass dimension, we must use the fractional calculus.
Smoothing of the microscopic characteristics over the 
physically infinitesimal volume, we transform the initial 
fractal medium into "fractional" continuous model
that uses the fractional integrals. 
The order of fractional integral is equal 
to the fractal mass dimension of the medium.

A more consistent approach to describe the fractal media
is connected with the mathematical definition of the integrals
on fractals. In Ref. \cite{RLWQ}, it was proven that integrals 
on net of fractals can be approximated by fractional integrals. 
In Ref. \cite{chaos} we proved that fractional integrals 
can be considered as integrals over the space with fractional 
dimension up to numerical factor. To prove this proposition, 
we use the well-known formulas of dimensional regularizations \cite{Col}.  

The fractional continuous models of fractal media \cite{PLA05}
can have a wide application. 
This is due in part to the relatively small numbers of parameters 
that define a random fractal medium of great complexity
and rich structure. The fractional continuous model \cite{PLA05} 
allows us to describe dynamics of a wide class of fractal media.  
Fractional integrals can be used to derive the fractional 
generalization of the equations for the fractal media.

Let us consider the region $W$ in three-dimensional 
Euclidean space $E^3$.
The volume of the region $W$ is denoted by $V(W)$.
If the region $W$ is a ball with the radius $R$,
then the volume $V(W_A)=(4/3)\pi R^3_A$ .
The mass of the region $W$ in the fractal medium is denoted 
by $M_D(W)$, where $D$ is a mass dimension of the medium. 

The properties of the fractal media like mass obeys a power law relation,
\be \label{MR-A} M(R) =kR^{D} , \quad (D<3),  \ee
where $M$ is the mass of fractal medium, $R$ is a box size (or a sphere radius),
and $D$ is a mass fractal dimension. 
The power law relation (\ref{MR-A}) can be naturally 
derived by using the fractional integral.
Let us consider the fractional generalization of the equation
\be \label{MW} M_3(W)=\int_W \rho({\bf r}) d^3 {\bf r} . \ee

Let us define the fractional integral in Euclidean space $E^3$ 
in the Riesz form \cite{SKM} by the equation
\be \label{ID} (I^{D}\rho)({\bf r}_0)=
\int_W \rho({\bf r}) dV_D , \ee
where $dV_D=c_3(D,r,r_0)d^3 {\bf r}$, and 
\[ c_3(D,r,r_0)=\frac{2^{3-D} \Gamma(3/2)}{\Gamma(D/2)} 
|{\bf r}-{\bf r}_0|^{D-3} ,  \quad
|{\bf r}-{\bf r}_0|=\sqrt{\sum^3_{k=1} (x_k-x_{k0})^2}. \]
The point ${\bf r}_0 \in W$ is the initial point of the fractional integral.
We will use the initial points in the integrals set to zero (${\bf r}_0=0$).
The numerical factor in Eq. (\ref{ID}) has this form in order to
derive the usual integral in the limit $D\rightarrow (3-0)$.
Note that the usual numerical factor
$\gamma^{-1}_3(D)={\Gamma(1/2)}/{2^D \pi^{3/2} \Gamma(D/2)}$,
which is used in Ref. \cite{SKM},  
leads to $\gamma^{-1}_3(3-0)= {\Gamma(1/2)}/{2^3 \pi^{3/2} \Gamma(3/2)}$ 
in the limit $D\rightarrow (3-0)$. 

Using notations (\ref{ID}), we can rewrite Eq. (\ref{MW})
in the form $M_3(W)=(I^{3}\rho)({\bf r}_0)$. 
Therefore the fractional generalization of this equation can be
defined in the form
\be \label{MWD}  M_D(W)=(I^D \rho)({\bf r}_0)=
 \frac{2^{3-D} \Gamma(3/2)}{\Gamma(D/2)}
\int_W \rho({\bf r}) |{\bf r}-{\bf r_0}|^{D-3} d^3 {\bf r} . \ee

If we consider the homogeneous fractal media 
($\rho({\bf r})=\rho_0=const$) and the ball region $W$, then
we have 
\[ M_D(W)= \rho_0 \frac{2^{3-D} \Gamma(3/2)}{\Gamma(D/2)} 
\int_W |{\bf R}|^{D-3} d^3 {\bf R} . \]
where ${\bf R}={\bf r}-{\bf r}_0$. 
Using the spherical coordinates, we get
\[ M_D(W)= \frac{\pi 2^{5-D} \Gamma(3/2)}{\Gamma(D/2)} 
\rho_0 \int_W R^{D-1} d R= 
\frac{2^{5-D} \pi \Gamma(3/2)}{D \Gamma(D/2)} \rho_0 R^{D} , \]
where $R=|{\bf R}|$. 
As the result, we have Eq. (\ref{MR-A}), 
where $k=[2^{5-D} \pi \Gamma(3/2) \rho_0]/[D \Gamma(D/2)]$
Therefore the fractal media with noninteger mass dimension $D$ can be
described by fractional integral of order $D$.



\end{document}